\documentclass{aa}  
\usepackage{graphicx}
\usepackage{multirow}
\newcommand{\bm}[1]{{\mbox{\boldmath $#1$}}}
\usepackage{color}

\usepackage{here}

\usepackage[colorlinks=true,citecolor=blue]{hyperref}
\usepackage{arydshln}
\usepackage{natbib}

\usepackage{comment}

\usepackage{booktabs}

\usepackage{adjustbox}

\usepackage{ulem}

\defcitealias{hanson2022}{HHS22}
\defcitealias{triana2022}{TGBR22}
\defcitealias{bhattacharya2022}{BH23}

\usepackage{txfonts}
\begin{document} 
   \title{
   Numerical study of non-toroidal inertial modes with \\ $l=m+1$
   radial vorticity in the Sun's convection zone
   }
   
   \titlerunning{Numerical study of non-toroidal inertial modes with $l=m+1$
   radial vorticity in the Sun's convection zone}

   \author{Yuto Bekki
          }

   \institute{Max-Planck-Institut f{\"u}r Sonnensystemforschung,
              Justus-von-Liebig-Weg 3, 37077 G{\"o}ttingen, Germany\\
              \email{\href{mailto:bekki@mps.mpg.de}{bekki@mps.mpg.de}}
             }

   \date{Received <-->; accepted <-->}


\abstract{
Various types of inertial modes have been observed and identified on the Sun, including the equatorial Rossby modes, critical-latitude modes, and high-latitude modes.
Recent observations further report a detection of equatorially-antisymmetric radial vorticity modes which propagate in a retrograde direction about three times faster than those of the equatorial Rossby modes when seen in the corotating frame with the Sun.
Here, we study the properties of these equatorially-antisymmetric vorticity modes using a realistic linear model of the Sun's convection zone.
We find that they are essentially non-toroidal, involving a substantial radial flow at the equator.
Thus, the background density stratification plays a critical role in determining their dispersion relation.
The solar differential rotation is also found to have a significant impact by introducing the viscous critical layers and confining the modes near the base of the convection zone.
Furthermore, we find that their propagation frequencies are strikingly sensitive to the background superadiabaticity $\delta$ because the buoyancy force acts as an additional restoring force for these non-toroidal modes.
The observed frequencies are compatible with the linear model only when the bulk of the convection zone is weakly subadiabatic ($-5\times 10^{-7} \lesssim \delta \lesssim -2.5\times 10^{-7}$).
Our result is consistent with but tighter than the constraint independently derived in a previous study ($\delta<2\times 10^{-7}$) employing the high-latitude inertial mode.
It is implied that, below the strongly superadiabatic near-surface layer, the bulk of the Sun's convection zone might be much closer to adiabatic than typically assumed and may even be weakly subadiabatic.
}


   \maketitle
%

\section{Introduction}

Inertial modes are global-scale low-frequency modes of oscillation in a rotating fluid whose restoring force is the Coriolis force \citep[e.g.,][]{greenspan1968}.
Recently, various kinds of inertial modes have been observed on the Sun.
These include the equatorial Rossby modes \citep[][]{loeptien2018,liang2019,proxauf2020,mandal2020,hathaway2020}, the critical-latitude modes, and the high-latitude modes \citep[][]{gizon2021}.
All of these modes propagate in a retrograde direction (opposite to solar rotation) when seen from the Carrington rotation frame.
It is expected that these inertial modes can be used to probe the interior of the Sun \citep[e.g.,][]{goddard2020,gizon2021,bekki2022a}.
In particular, using the baroclinically-unstable $m=1$ high-latitude mode, \citet[][]{gizon2021} derived the observational constraint of the mean superadiabaticity $\delta$, one of the most important unknown parameters in the Sun's convection zone.
They deduced $\delta < 2\times 10^{-7}$, implying that the Sun's convection zone is closer to adiabatic than typically assumed.


Recently, \citet[][hereafter \citetalias{hanson2022}]{hanson2022} have reported to detect another family of modes near the surface of the Sun, i.e., retrograde-propagating modes of equatorially antisymmetric radial vorticity $\zeta_{r}$.
They can be most clearly seen in the $l=m+1$ component of the radial vorticity power spectrum, where $l$ is the spherical harmonic degree and $m$ is the azimuthal order.
\citetalias[][]{hanson2022} reported that these modes propagate in a retrograde direction about three times faster than the equatorial Rossby modes with the same azimuthal order $m$, which cannot be explained by the classical Rossby modes.
In this paper, we call them HHS22 modes hereafter.

A possible identification of the HHS22 modes was first provided by \citet[][hereafter \citetalias{triana2022}]{triana2022} using a linear model of rotating fluid in a spherical shell.
They are identified as a particular class of non-toroidal inertial modes which involve substantial radial motions in the middle convection zone.
The computed linear dispersion relation of the HHS22 modes shows a good agreement with their observed frequencies.
However, the linear model of \citetalias[][]{triana2022} was highly simplified, e.g., the model assumes an incompressible (uniform density) fluid and uniformly-rotating convection zone, which are not appropriate in the Sun.

A similar mode identification was later reported by \citet[][hereafter \citetalias{bhattacharya2022}]{bhattacharya2022} in which the solar-like background stratification was taken into account using the anelastic approximation.
However, the latitudinal differential rotation of the Sun was still omitted in their model for simplicity, which is known to have a substantial impact on the properties of inertial modes \citep[e.g.,][]{baruteau2013,gizon2021,bekki2022a,fournier2022,philidet2023,bhattacharya2023b}.
\citetalias[][]{bhattacharya2022} found that the computed frequencies of the HHS22 modes are lower than the observed ones by about $100$ nHz, in contrast to the match found by \citetalias[][]{triana2022}.
It is necessary to understand the origin of this discrepancy and to investigate the missing physics to properly reproduce the observed features of the HHS22 modes.

In this paper, we study the properties of the HHS22 modes using a more realistic linear model of the Sun's convection zone \citep[][]{bekki2022a} which takes into account both the solar density stratification and the solar differential rotation determined by global helioseismology \citep[][]{larson2018}.
We will then show that their frequencies are strongly affected by the superadiabaticity $\delta$ in the bulk convection zone.

Recently, the solar inertial modes are also studied using fully-nonlinear simulations of the rotating convection.
\citet[][]{bekki2022b} have mostly focused on the equatorial Rossby modes and the columnar convective (also known as thermal Rossby) modes.
\citet[][]{matilsky2022} have implied that the equatorial Rossby modes might contribute to the confinement of the solar tachocline via dynamo action in the radiative interior. 
However, the HHS22 modes have never been studied yet.
In Appendix of this paper, we will also show that the HHS22 modes can be found to exist in the fully-nonlinear simulations.

The organization of the paper is as follows.
In \S~\ref{sec:linana_method}, our linear eigenmode solver is explained.
The effects of the background density stratification and the solar differential rotation on the HHS22 modes are examined in \S~\ref{sec:comparison}.
The effects of turbulent viscosity is briefly discussed in \S~\ref{sec:linana_viscosity}.
We then show how the solar observations can be used to infer the superadiabaticity in the bulk of the convection zone in \S~\ref{sec:linana_delta}.
In Appendix~\ref{app:nonlinear}, we further report that the HHS22 modes are found to exist in the fully-nonlinear simulations of rotating convection.
The conclusions are summarized in \S~\ref{sec:summary}.

\begin{table}[]
 \begin{center} 
\caption{Summary of the linear analysis model setups.}
\small
\vspace{-0.3\baselineskip}
\begin{tabular}{cccccccccc} 
\toprule
\toprule
  \renewcommand{\arraystretch}{1.8}
Model & & Stratification & Differential rotation &  \\
 \midrule
1 &...& No (incompressible) & No (uniform rotation)    \\
2 &...& Yes (solar-like)    & No (uniform rotation)    \\
3 &...& Yes (solar-like)    &  Yes (solar-like)   \\
\bottomrule
\end{tabular}
\label{table:lin}
\end{center}
\vspace{-1.0\baselineskip}
\tablefoot{
Model 1 assumes the incompressible fluid and uniform rotation, corresponding to that of \citetalias[][]{triana2022}.
In model 2, the solar-like density stratification is taken into account but the uniform rotation is still assumed, which is similar to that of \citetalias[][]{bhattacharya2022}.
In model 3, we consider both the background density stratification and the solar differential rotation determined by global helioseismology \citep[][]{larson2018}.
} 
\end{table}


\begin{figure}
\begin{center}
\includegraphics[width=0.925\linewidth]{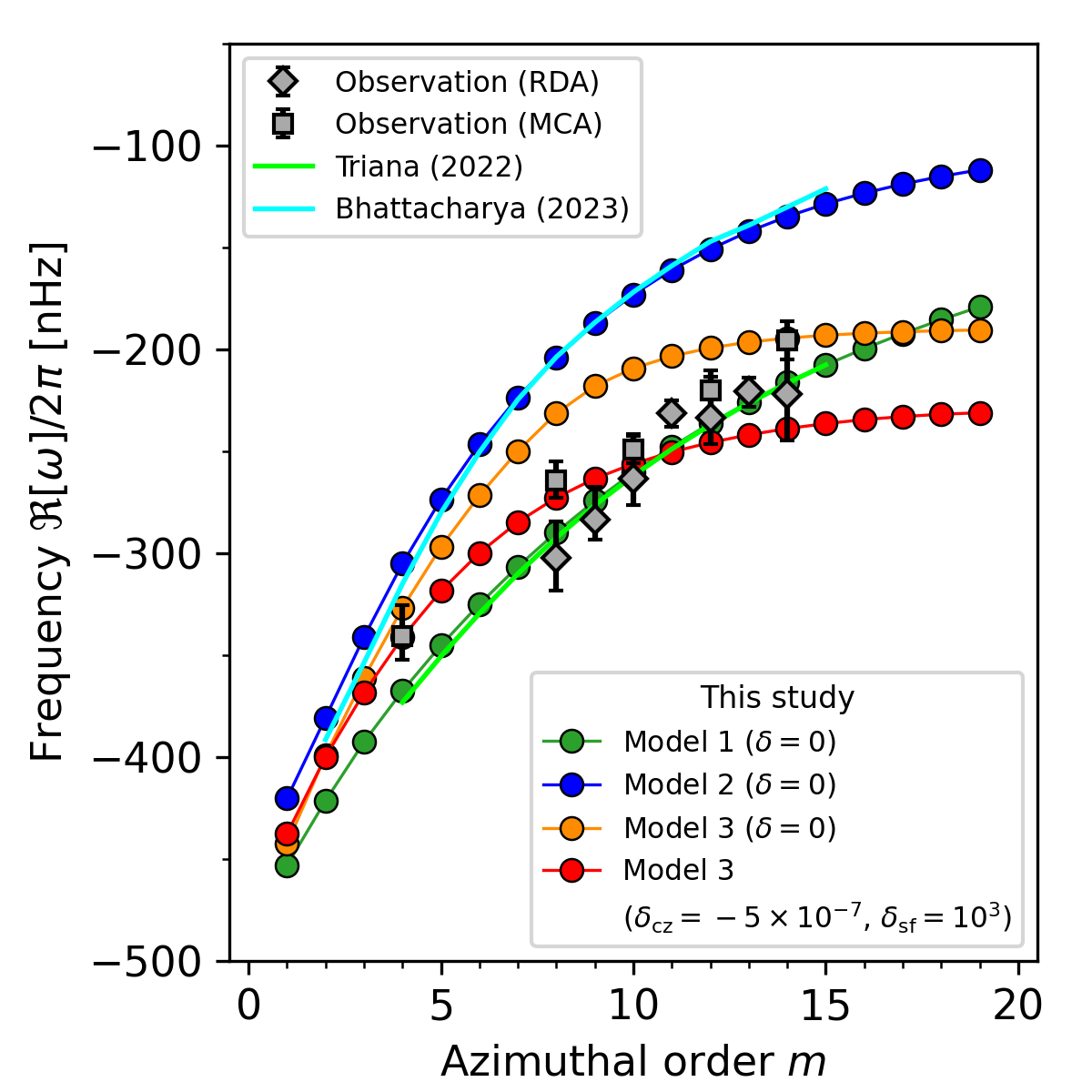}
\caption{
Dispersion relation of the HHS22 modes obtained from the linear analysis for $1 \leq m \leq 19$.
Green points represent the results from our model 1 where we assume an incompressible (constant density) fluid and an uniform rotation.
Blue points represent the results from our model 2 where the solar background stratification is included but the uniform rotation is still assumed.
Orange points represent the results from our model 3 where both the solar stratification and the solar differential rotation are included.
Red points also show the results from model 3 but with a weakly subadiabatic bulk convection zone ($\delta_{\mathrm{cz}}=-5\times 10^{-7}$) and a strongly superadiabatic near-surface layer ($\delta_{\mathrm{sf}}=10^{-3}$).
Lime and cyan solid curves show the results from \citetalias[][]{triana2022} and from \citetalias[][]{bhattacharya2022}, respectively.
For comparison, we also show the observed frequencies of the HHS22 modes reported in \citetalias{hanson2022} where the gray diamonds and squares denote the measurements with ring-diagram analysis (RDA) and mode-coupling analysis (MCA), respectively.
All the frequencies are measured in the Carrington frame rotating at $\Omega_{0}/2\pi=456$ nHz.
}
\label{fig:dispersion}
\end{center}
\end{figure}
\begin{figure*}
\begin{center}
\includegraphics[width=0.85\linewidth]{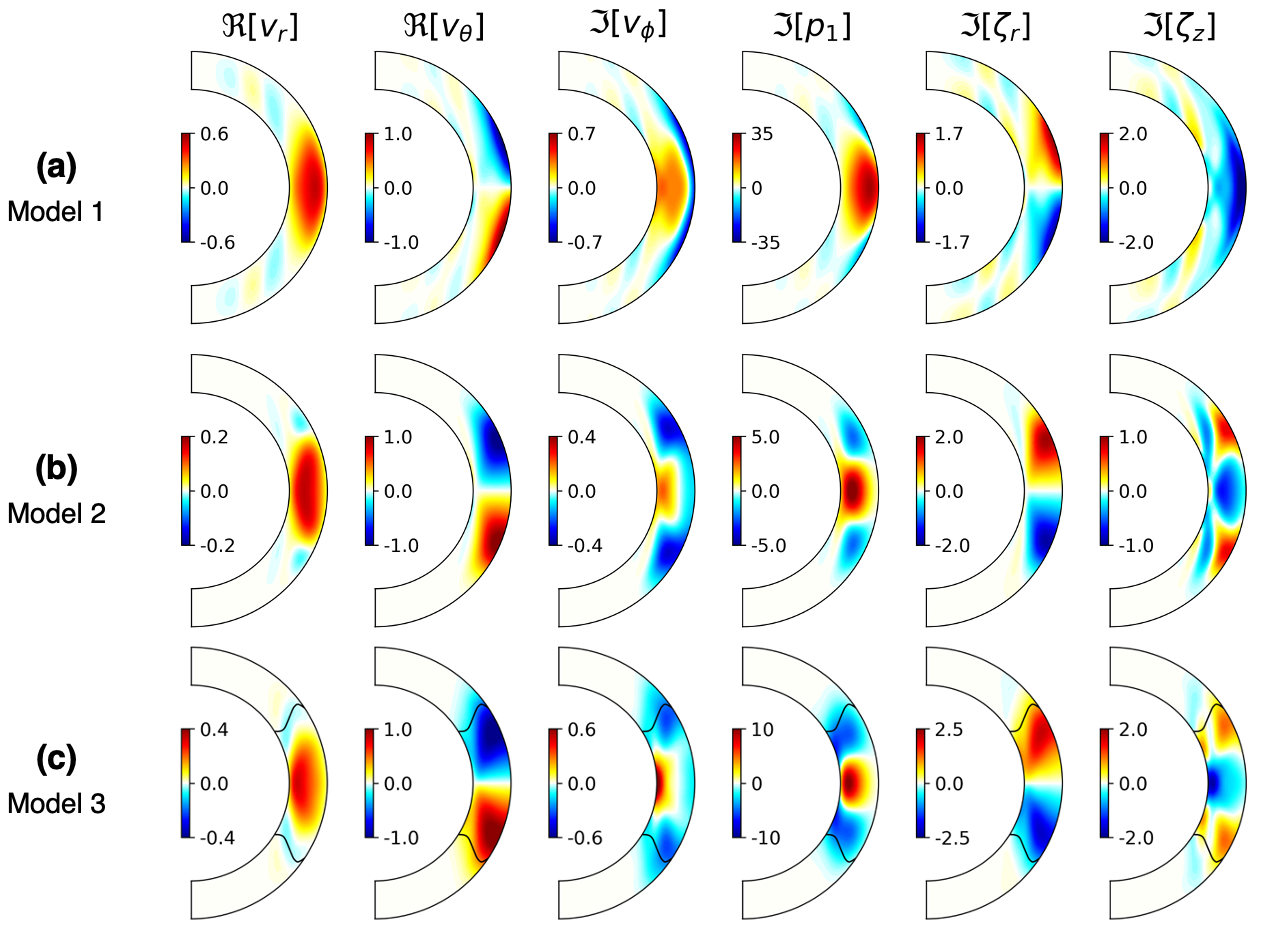}
\caption{
Meridional cuts of the eigenfunctions of the $m=10$ HHS22 mode obtained from the linear analysis.
The velocity, pressure, and vorticity eigenfunctions are expressed as $\bm{v}(r,\theta) \exp{[i(m\phi-\omega t)]}$, $p_{1}(r,\theta) \exp{[i(m\phi-\omega t)]}$, and $\bm{\zeta}(r,\theta) \exp{[i(m\phi-\omega t)]}$, and the solutions are shown in the meridional plane at $t=0$ and $\phi=0$.
The units of the colorbars are m~s$^{-1}$ for three velocity components, $10^{4}$ dyn~cm$^{-2}$ for the pressure perturbation, and $10^{-8}$ s$^{-1}$ for the vorticity components.
The eigenfunctions are normalized such that the maximum of $v_{\theta}$ is $1.0$~m~s$^{-1}$.
Panels (a--c) show the results from models 1--3, respectively.
Note that the background stratification is adiabatic in all cases.
In panel (c), the black solid curves show the locations of the critical latitudes.
}
\label{fig:eigenfunc_lin}
\end{center}
\end{figure*}


\section{Method: linear eigenmode analysis} \label{sec:linana_method}

We use the code developed by \citet[][]{bekki2022a}, which numerically solves the linear eigenvalue problem for a rotating fluid in a spherical shell $0.71R_{\odot} < r < 0.985R_{\odot}$.
Here, $R_{\odot}$ is the solar radius.
The model is hydrodynamic, i.e., the effects of magnetic field are ignored for simplicity.
The eigenvalue equations are solved for azimuthal orders $1 \leq m \leq 19$.
To study the HSS22 modes, we seek for the retrograde-propagating modes ($\Re[\omega]<0$) whose eigenfunctions satisfy the following criteria:
\begin{itemize}
    \item The eigenfunction of radial velocity $v_{r}$ is dominantly $l=m$ in the middle convection zone, and has no radial node at the equator.
    \item The eigenfunction of latitudinal velocity $v_{\theta}$ is dominantly $l=m+1$ at the surface, and has no radial node at low latitudes.
    \item The eigenfunction of latitudinal velocity $v_{\phi}$ is dominantly $l=m$ or $l=m+2$ at the surface.
\end{itemize}
When the above criteria are satisfied by multiple eigenmodes, we select the least-damped mode (with largest growth rate $\Im[\omega]$) among them.
For further details, see \citet[][\S~2]{bekki2022a}.

In this paper, we first carry out three sets of linear eigenmode calculations with different model setups.
The first setup (model 1) consists of incompressible fluid and uniform rotation, which were assumed in the previous study of \citetalias[][]{triana2022}.
In the second setup (model 2), we still assume the uniform rotation but include the solar background density stratification, which is similar to the numerical setup of \citetalias[][]{bhattacharya2022}.
The third setup (model 3) finally takes into account both the realistic solar density stratification and the solar differential rotation determined by global helioseismology \citep{larson2018}.
These are summarized in Table~\ref{table:lin}.

\section{Results} \label{sec:linana_results}

\subsection{Comparison with previous studies} \label{sec:comparison}

In this section, we compare the results from models 1--3.
In all cases, we include the spatially constant viscous diffusivity of $\nu=10^{12}$ cm$^{2}$~s$^{-1}$.
For simplicity, we assume that the background is purely adiabatic, i.e., there is no entropy variation neither in the radial nor the latitudinal directions.

\subsubsection{Effects of density stratification} \label{sec:linana_stratification}

Figure~\ref{fig:dispersion} shows the linear dispersion relations of the HHS22 modes computed from the models 1-3, measured in the Carrington frame rotating at $\Omega_{0}/2\pi=456.0$ nHz.
We find that the dispersion relation from the model 1 matches almost exactly to that of \citetalias{triana2022}; the differences are found to be less than $1.5\%$.
It is confirmed that the observed frequencies of the HHS22 modes can be nicely fitted by the dispersion relation of our model 1 where the over-simplifying assumptions are used.
Interestingly, however, when the solar density stratification is taken into account, the dispersion relation of the HHS22 modes deviates from the observations:
In our model 2, the frequencies become much lower than the observed ones (by about $100$ nHz at $m=10$).
These frequencies are consistent with the results reported in \citetalias[][]{bhattacharya2022} with the differences on the order of few percent.
The small discrepancy can be attributed to the differences in the model setups such as the different radial profiles of the turbulent diffusivities and the inclusion of the radiative interior below the convection zone.

\begin{figure*}
\begin{center}
\includegraphics[width=0.875\linewidth]{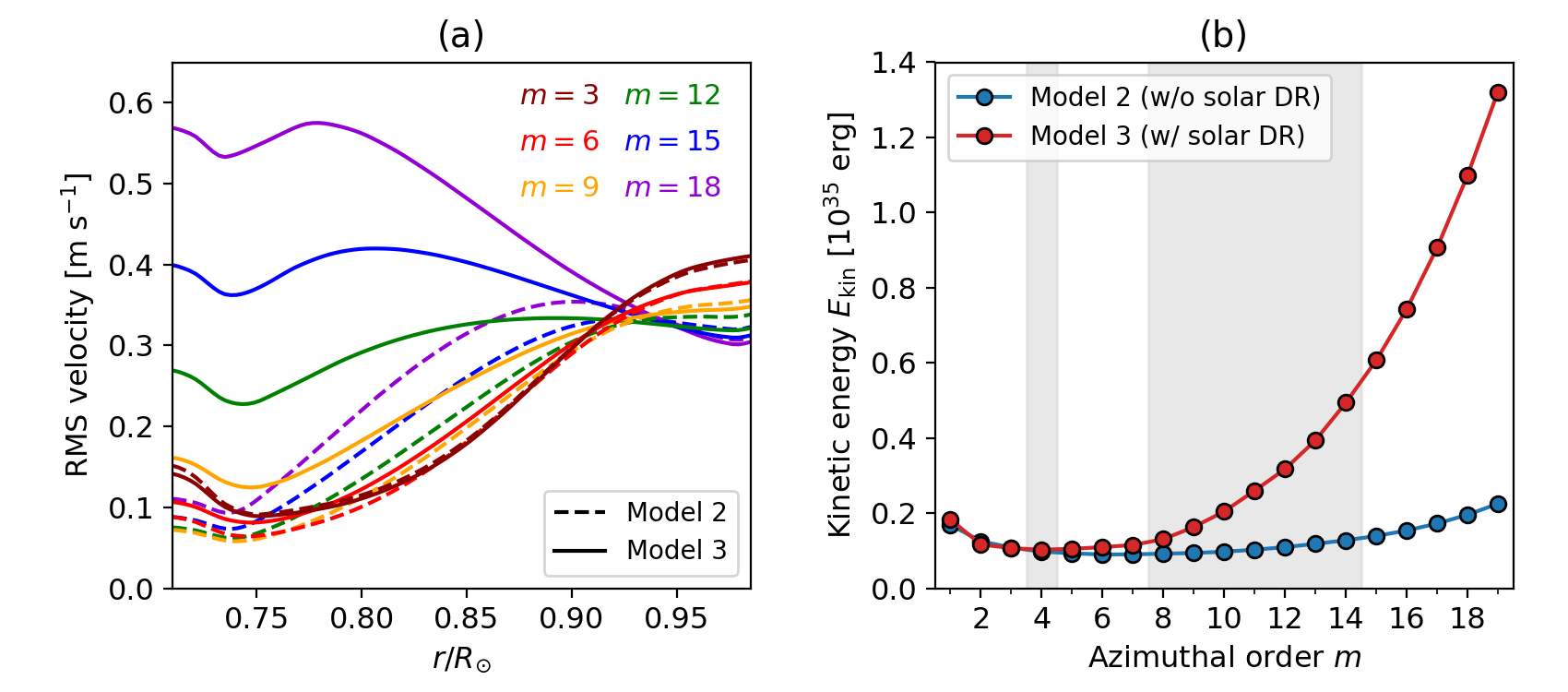}
\caption{
(a) Root-mean-square (RMS) amplitudes of the velocity eigenfunctions of the HHS22 modes from the linear analysis, where the average is taken over the spherical surfaces.
Dashed and solid curves show the results from model 2 (without differential rotation) and model 3 (with differential rotation).
Different colors show different azimuthal orders $m$.
The eigenfunctions are normalized such that the maximum horizontal velocity at $r=0.985R_{\odot}$ is $1.0$~m~s$^{-1}$.
(b) Spectra of volume-integrated kinetic energy of the HHS22 modes.
The blue and red points denote those from model 2 and 3, respectively.
}
\label{fig:vrms_Ekin}
\end{center}
\end{figure*}

Figure~\ref{fig:eigenfunc_lin} shows the meridional eigenfunctions of the HHS22 mode at $m=10$.
All the eigenfunctions are normalized such that the maximum of $v_{\theta}$ is $1.0$~m~s$^{-1}$ at the surface, as suggested by the observation \citepalias{hanson2022}.
Those of model 1 (Fig.~\ref{fig:eigenfunc_lin}a) agree with the results of \citetalias[][]{triana2022} (see Fig.~4 top panels therein).
As already reported, the HHS22 modes have a radial vorticity $\zeta_{r}$ which is north-south antisymmetric across the equator.
At the equator, it is seen that the latitudinal diverging (converging) motion at the surface involves a substantial radial upflow (downflow) in the middle convection zone.
This clearly shows that the HHS22 modes are not toroidal at all.
Therefore, their mode properties are different from the $l=m+1$ classical Rossby modes which are quasi-toroidal (even though their surface eigenfunctions look similar to those of $l=m+1$ classical Rossby modes).

To further investigate the consequences of the non-toroidalness, we show the eigenfunctions of $z$-vorticity $\zeta_{z}$ in the rightmost panels of Fig.~\ref{fig:eigenfunc_lin}, where $z$ denotes a direction parallel to the rotational axis.
It is seen that the HHS22 mode involves a north-south symmetric $z$-vortical motion near the equator, in addition to the north-south antisymmetric $r$-vortical motion.
We note that the amplitude of $\zeta_{z}$ is comparable to that of $\zeta_{r}$.
This $z$-vortical motion invokes additional $\beta$-effects, i.e., the topographic $\beta$-effect originating from the spherical curvature \citep[e.g.,][]{busse2002} and the compressional $\beta$-effect originating from the background density stratification \citep[e.g.,][]{glatzmaier2009,verhoeven2014,gastine2014,bekki2022a}.
Outside the tangential cylinder of the Sun's convection zone, it is known that the compressional $\beta$-effect dominates over the topographic $\beta$-effect \citep[][see Fig.~3.32 therein]{bekki_thesis}.
At the equator, the $z$-vorticity equation can be expressed as
\begin{eqnarray}
    && \frac{\partial \zeta_{z}}{\partial t}=\beta_{\mathrm{comp}}v_{r} + [ ... ], 
    \ \ \ \mathrm{with} \ \ \ \beta_{\mathrm{comp}}=-\frac{2\Omega_{0}}{H_{\rho}},
\end{eqnarray}
where $H_{\rho}$ denotes the density scale height of the background, and we only retain the term related to the compressional $\beta$-effect.
When the simplifying assumption of incompressible fluid is used in model 1, the compressional $\beta$-effect is ignored ($\beta_{\mathrm{comp}}=0$).
However, when the density stratification is included in model 2, a negative $\beta_{\mathrm{comp}}$ promotes a prograde propagation which acts against the retrograde propagation of the modes \citep[][]{glatzmaier1981}.
This decreases the retrograde propagation frequencies of the HHS22 modes compared to the model 1, and consequently, the dispersion relation deviates from the observations.
Our study suggests that a great agreement reported by \citetalias[][]{triana2022} is largely due to the unrealistic assumption of incompressible fluid in the Sun's convection zone, which ignores the compressional $\beta$-effect.

\subsubsection{Effects of solar differential rotation} \label{sec:linana_diffrot}

It is shown in Fig.~\ref{fig:dispersion} that the inclusion of solar differential rotation changes the dispersion relation of the HHS22 modes (from model 2 to model 3).
The significant modification occurs at higher $m$ where the retrograde propagation frequencies become larger than the observed values by about $50-80$ nHz.
This is a direct consequence of the Doppler frequency shift by the differential rotation.

The eigenfunctions of the $m=10$ HHS22 mode from the model 3 are shown in Fig.~\ref{fig:eigenfunc_lin}c.
It is known that the inclusion of latitudinal differential rotation gives rise to critical latitudes where the phase speed of the mode becomes equal to the local differential rotation speed \citep[][]{baruteau2013,gizon2020,bekki2022a,fournier2022,philidet2023}.
The locations of the critical latitudes are denoted by black solid curves.
Compared to those from the model 2, the mode eigenfunctions from the model 3 are distorted by the existence of critical layers.

To better see the impact of the critical layers, we show the radial profiles of the root-mean-square (RMS) velocity eigenfunctions of the HHS22 modes for different azimuthal orders $m$ in Fig.~\ref{fig:vrms_Ekin}a.
For small azimuthal orders $m \ ( \lesssim 6)$, the radial eigenfunctions from models 2 and 3 are similar:
In both cases, the RMS velocity increases with radius and peaks at the surface.
However, as $m$ increases, the velocity eigenfunctions from the model 3 are more and more confined into the lower convection zone, leading to a significant deviation from those of model 2.
This is because the HHS22 modes tend to be more and more localized around the critical layers.
For sufficiently large $m \ (\gtrsim 10)$, the critical layers exist near the base of the convection zone at the equator.
Figure~\ref{fig:vrms_Ekin}b further compares the volume-integrated kinetic energy of the HHS22 modes between the model 2 and model 3 where the maximum horizontal velocity amplitudes are fixed to $1.0$~m~s$^{-1}$ at the surface in both cases.
It is shown that, as a consequence of the mode confinement near the base, the HHS22 modes have much more kinetic energy in model 3 than in model 2.
Therefore, our study suggests that the solar differential rotation needs to be properly taken into account in order to evaluate the dynamical importance of the HHS22 modes in the Sun's convection zone.
In the remaining part of the paper, we only consider the most realistic model 3 which takes into account both the solar density stratification and the solar differential rotation.

\begin{figure*}
\begin{center}
\includegraphics[width=0.875\linewidth]{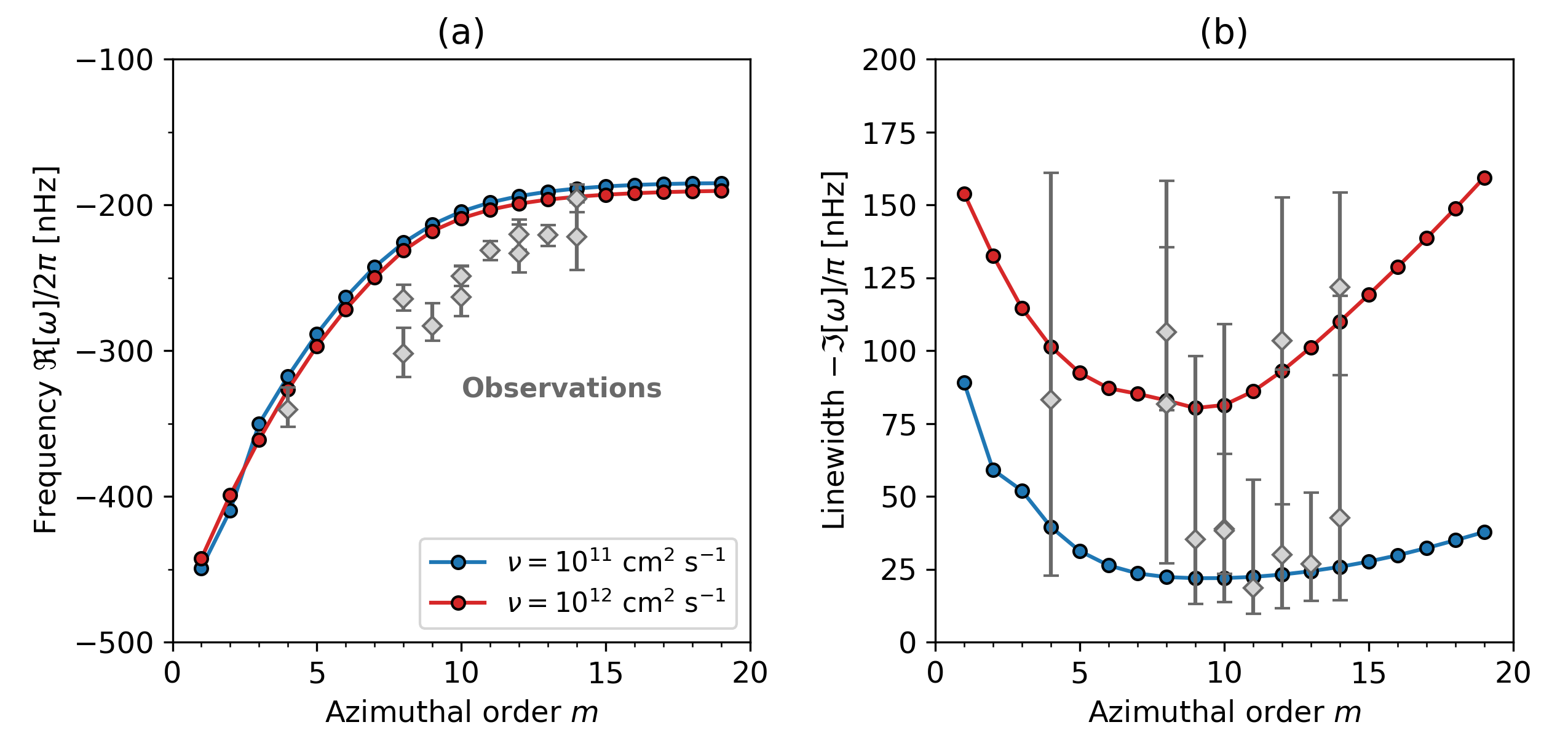}
\caption{
(a) Propagation frequencies and (b) linewidths of the HHS22 modes obtained from the linear calculations with two different values of turbulent viscosity $\nu$.
Blue and red points show the results with spatially uniform viscosity of $\nu=10^{11}$~cm$^{2}$~s$^{-1}$ and for $10^{12}$~cm$^{2}$~s$^{-1}$, respectively.
The model 3 is used (with solar density stratification and solar differential rotation) and the background is assumed to be adiabatic ($\delta=0$).
The observed values reported by \citetalias{hanson2022} are shown by gray diamonds.
}
\label{fig:visc}
\end{center}
\end{figure*}

\subsection{Dependence on turbulent viscosity} \label{sec:linana_viscosity}

In this section, the dependence of the HHS22 modes on the turbulent viscosity $\nu$ is examined.
Here, we use the model 3 with the adiabatic background, i.e., $\delta=0$ throughout the convection zone.
Figure~\ref{fig:visc} compares the eigenfequencies of the HHS22 modes obtained for two different values of turbulent viscosity, $\nu=10^{11}$ and $10^{12}$~cm$^{2}$~s$^{-1}$.
For simplicity, the viscosity $\nu$ is assumed to be spatially constant throughout the convection zone.
Although the mode linewidths are decreased by about $60-100$~nHz with the decrease of $\nu$, the dispersion relation of the HHS22 modes is only marginally affected (difference of less than $3\%$).
Regardless of the values of turbulent viscosity used, the discrepancies between our linear model and the observations remain in the propagation frequencies of the HHS22 modes at $8 \leq m \leq 14$.

\subsection{Dependence on superadiabaticity $\delta$} \label{sec:linana_delta}

Next, we investigate the effects of the non-adiabatic stratification in the Sun's convection zone on the HHS22 modes.
A deviation from the adiabatic stratification is measured by the superadiabaticity $\delta =\nabla -\nabla_{\mathrm{ad}}$ where $\nabla = d\ln{T}/d\ln{p}$.
The superadiabaticity in the Sun's convection zone is typically estimated to be very small, $\delta \approx \mathcal{O}(10^{-6})$ \citep[][]{ossendrijver2003}, except for a very thin near-surface layer where the solar granulation is vigorously driven by the strong surface radiative cooling \citep[e.g.,][]{nordlund2009}.
According to the solar standard model S \citep[][]{christensen1996,stix2002}, $\delta$ is expected to increase up to $\mathcal{O}(10^{-3})$ at $r \approx 0.99R_{\odot}$ and even further up to $\mathcal{O}(10^{-1})$ at the photosphere.

In this section, we carry out a set of linear calculations with varying superadiabaticity $\delta$.
The most realistic model 3 is used with the turbulent viscousity of $\nu=10^{12}$~cm$^{2}$~s$^{-1}$.
We use the following radial profiles of $\delta(r)$ which mimics a sharp transition from the close-to-adiabatic bulk convection zone to the strongly superadiabatic surface,
\begin{eqnarray}
&& \delta(r)=\delta_{\mathrm{cz}} + (\delta_{\mathrm{sf}}-\delta_{\mathrm{cz}}) \exp{\left[-\left(\frac{r-r_{\mathrm{max}}}{d_{\mathrm{sf}}} \right)^{2} \right]}, \label{eq:deltar}
\end{eqnarray}
where $\delta_{\mathrm{cz}}$ and $\delta_{\mathrm{sf}}$ denote the values of superadiabaticity in the bulk of the convection zone and near the top surface, respectively.
We use the values $d_{\mathrm{sf}}=0.015R_{\odot}$ and $r_{\mathrm{max}}=0.985R_{\odot}$.
Therefore, the strong superadiabaticities are localized in a thin layer near the top boundary of our numerical domain.

\begin{figure*}
\begin{center}
\includegraphics[width=0.95\linewidth]{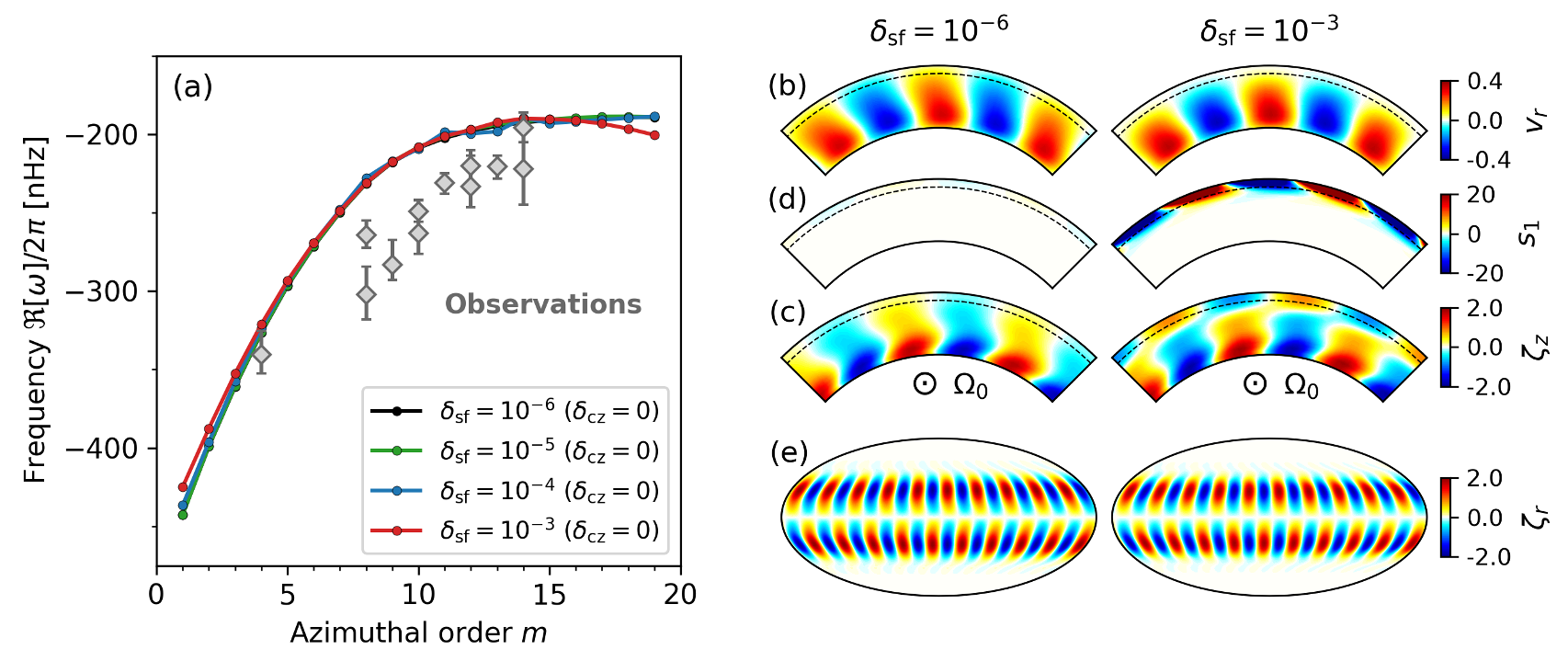}
\caption{
(a) Dispersion relations of the HHS22 modes computed for different values of superadiabaticity near the surface $\delta_{\mathrm{sf}}$.
The bulk of the convection zone is assumed to be adiabatic, $\delta_{\mathrm{cz}}=0$. 
(b--e) Eigenfunctions of the $m=10$ HHS22 mode computed for $\delta_{\mathrm{sf}}=10^{-6}$ (left column) and for $\delta_{\mathrm{sf}}=10^{-3}$ (right column).
Panels (b--d) show cuts of radial velocity $v_{r}$ (in m~s$^{-1}$), $z$-vorticity $\zeta_{z}$ (in $10^{-8}$~s$^{-1}$), and entropy perturbation $s_{1}$ (in erg~g$^{-1}$~K$^{-1}$) in the equatorial plane (along the rotational axis) seen from the north pole.
The black dashed curves denote the height $r=0.95R_{\odot}$, above which the strongly superadiabatic layer is located.
(e) Mollweide projection of the radial vorticity eigenfunction $\zeta_{r}$ at the top boundary $r=0.985R_{\odot}$ (in $10^{-8}$~s$^{-1}$).
All the eigenfunctions are normalized in the same way as in Fig.~\ref{fig:eigenfunc_lin}.
}
\label{fig:ad_supsf}
\end{center}
\end{figure*}
\begin{figure*}
\begin{center}
\includegraphics[width=0.95\linewidth]{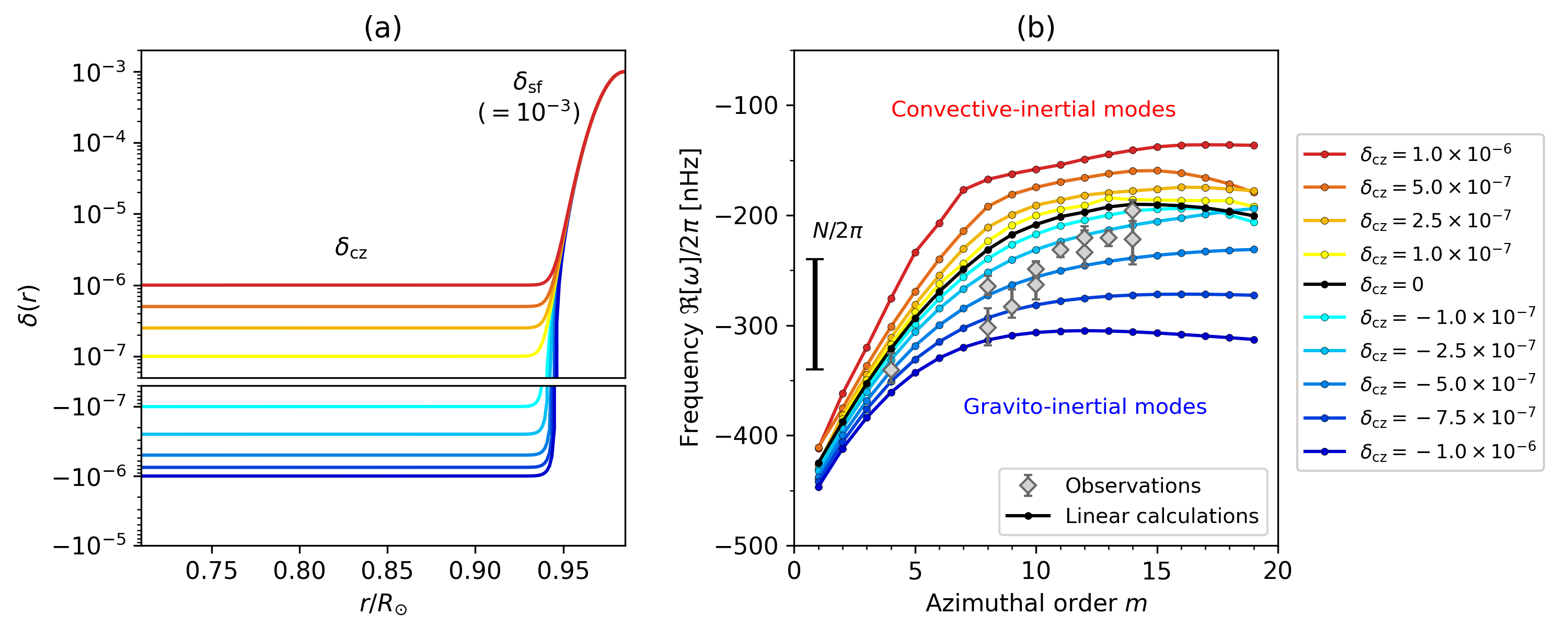}
\includegraphics[width=0.95\linewidth]{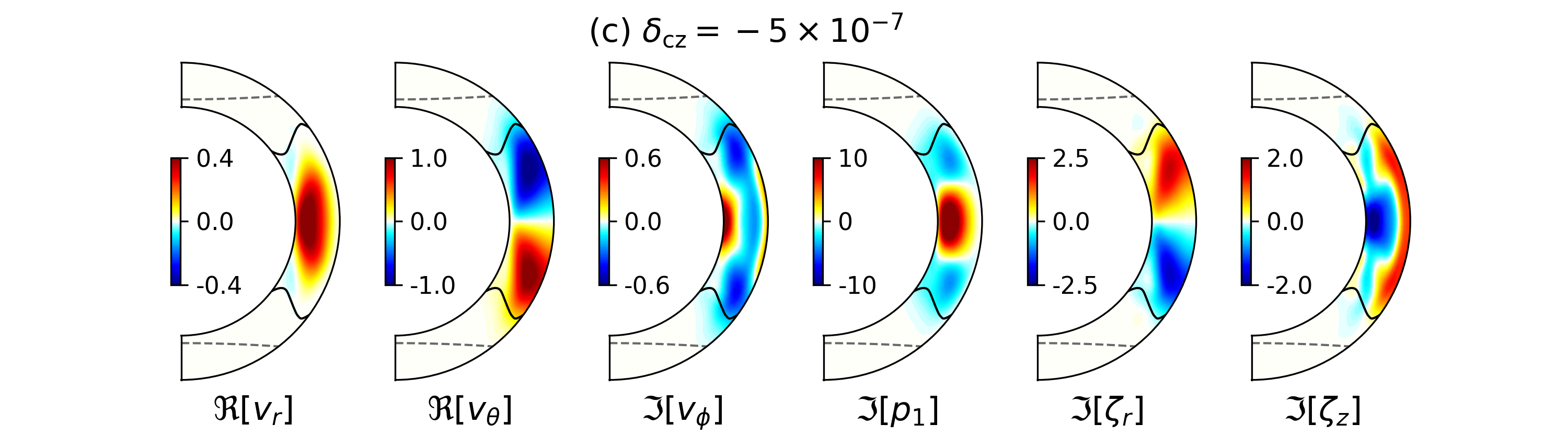}
\caption{
(a) Radial profiles of the superadiabaticity $\delta(r)$ defined by Eq.~(\ref{eq:deltar}) for different values of the bulk superadiabaticity $\delta_{\mathrm{cz}}$.
The superadiabaticity in the near-surface layer is fixed to $\delta_{\mathrm{sf}}=10^{-3}$.
(b) Dispersion relations of the HHS22 modes computed for different superadiabaticity profiles as shown in panel (a).
The observed frequencies reported by \citetalias{hanson2022} are denoted by gray diamonds.
(c) Meridional eigenfunctions from the case with weakly subadiabatic bulk convection zone ($\delta_{\mathrm{cz}} = -5 \times 10^{-7}$).
Black solid and gray dashed curves show the locations of the critical latitudes (by differential rotation) and the turning surfaces (by subadiabatic stratification), respectively.
}
\label{fig:disp_growth_supsf}
\end{center}
\end{figure*}

\subsubsection{Impact of near-surface superadiabaticity} \label{sec:delta_sf}

Figure~\ref{fig:ad_supsf}a shows the linear dispersion relations of the HHS22 modes computed for different values of the near-surface superadiabaticity $\delta_{\mathrm{sf}}$.
In all cases, the bulk of the convection zone (below $0.95R_{\odot}$) is fixed to be purely adiabatic, i.e., $\delta_{\mathrm{cz}}=0$.
Within the parameter range investigated here, the frequencies of the HHS22 modes are shown to be almost independent of $\delta_{\mathrm{sf}}$. 

Figure~\ref{fig:ad_supsf}b compares the eigenfunctions of the $m=10$ HHS22 mode from the two representative cases with $\delta_{\mathrm{sf}}=10^{-6}$ and with $\delta_{\mathrm{sf}}=10^{-3}$.
Figs.~\ref{fig:ad_supsf}b--d show the eigenfunctions of radial velocity $v_{r}$, $z$-vorticity $\zeta_{r}$, and entropy perturbation $s_{1}$ in the equatorial plane, respectively.
The radial velocity $v_{r}$ is almost unchanged being concentrated near the base of the convection zone.
However, the strong entropy perturbation $s_{1}$ is generated in the near-surface superadiabatic layer as $\delta_{\mathrm{sf}}$ increases (Fig.~\ref{fig:ad_supsf}c).
The buoyancy force associated with this entropy perturbation drives the $z$-vortical motion in this near-surface layer (Fig.~\ref{fig:ad_supsf}d).
Nonetheless, it is shown that the overall structure of the mode eigenfunctions in the bulk of the convection zone (below $0.95R_{\odot}$) is only little affected by the inclusion of the strongly superadiabatic near-surface layer.
In fact, the surface eigenfunction of radial vorticity $\zeta_{r}$ remains unchanged by the increase of $\delta_{\mathrm{sf}}$ up to $10^{-3}$ (Fig.~\ref{fig:ad_supsf}e).

We must note that, in the real Sun, $\delta$ is expected to further increase above $0.985R_{\odot}$ up to $\mathcal{O}(10^{-1})$ near the photosphere \citep[e.g.,][]{christensen1996,stix2002}.
An inclusion of this realistic photosphere may have a non-negligible impact on the HHS22 modes.
However, resolving this very thin surface layer is numerically challenging and thus beyond the scope of this paper.

\subsubsection{Impact of superadiabaticity in the bulk convection zone} \label{sec:delta_cz}

Next, we vary the superadiabaticity in the bulk of the convection zone $\delta_{\mathrm{cz}}$ within a range of $\pm 10^{-6}$ while fixing a value for the near-surface superadiabaticity $\delta_{\mathrm{sf}}=10^{-3}$. 
The radial profiles of $\delta(r)$ used in this study are shown in Fig.~\ref{fig:disp_growth_supsf}a.

Figure~\ref{fig:disp_growth_supsf}b manifests a striking sensitivity of the dispersion relation of the HHS22 modes to a tiny change in $\delta_{\mathrm{cz}}$.
It is found that, as the stratification becomes more superadiabatic (subadiabatic), their dispersion relation shifts towards more positive (negative) direction in frequency, i.e., the HHS22 modes propagate in a retrograde direction with slower (faster) phase speed when $\delta_{\mathrm{cz}}>0 \ (<0)$.
This frequency shift can be understood by considering whether the buoyancy force associated with the radial motion acts as an additional restoring force or the opposite.
The similar behavior is already reported and discussed in the case of columnar convective (thermal Rossby) modes by \citet[][see \S~5.]{bekki2022a}, in which their prograde propagation frequencies become slower (faster) when $\delta>0 \ (<0)$.
It is shown in Fig.~\ref{fig:disp_growth_supsf}b that the observed frequencies of the HHS22 modes can be nicely fitted by the linear dispersion relation with weakly subadiabatic bulk convection zone $-5\times 10^{-7} \lesssim \delta_{\mathrm{cz}} \lesssim -2.5\times 10^{-7}$.
This is within the range of the observational constraint of $\delta$ independently derived by \citet[][]{gizon2021} based on the $m=1$ high-latitude inertial mode.

For the range of subadiabaticity $-5\times 10^{-7} \lesssim \delta_{\mathrm{cz}} \lesssim -2.5\times 10^{-7}$, the Brunt-Väisälä frequency $N=\sqrt{g|\delta_{\mathrm{cz}}|/H_{p}}$ is estimated to be $N/2\pi \approx 240-340$~nHz (with $g \approx 520$~m~s$^{-1}$ and $H_{p} \approx 57$~Mm near the base of the convection zone), which is comparable to the mode frequencies at $8 \leq m \leq 14$.
Therefore, the HHS22 modes are expected to behave as gravito-inertial modes in which both Coriolis and buoyancy forces act as restoring forces. 
Some gravito-inertial modes are known to be trapped by turning surfaces \citep[][]{friedlander1982,dintrans1999,mirouh2016}. 
The eigenfunctions of the $m=10$ HHS22 mode are shown in Fig.~\ref{fig:disp_growth_supsf}c in the case of $\delta_{\mathrm{cz}}=-5\times 10^{-7}$.
We find that the turning surfaces (denoted by gray dashed curves) are located at higher latitudes than the critical latitudes of the differential rotation.
Therefore, the turning surfaces only play a limited role in trapping the HHS22 modes which are already strongly confined into the equatorial region by the critical latitudes.
In the bulk of the convection zone, therefore, no significant difference can be seen in the mode eigenfunctions from the adiabatic case (Fig.~\ref{fig:eigenfunc_lin}c).
Near the surface, by contrast, strong $z$-vortical motions are apparent as a consequence of the strong superadiabaticity.

\section{Summary and Discussion} \label{sec:summary}

In this study, we investigated the properties of the $l=m+1$ radial vorticity modes recently observed by \citetalias[][]{hanson2022} (we call them HHS22 modes in this paper). 
We used a linear eigenmode solver of the Sun's convection zone developed by \citet[][]{bekki2022a}.
Our model can successfully reproduce the previous results of \citetalias[][]{triana2022} and \citetalias[][]{bhattacharya2022} when the simplifying assumptions are used.
We found that, in contrast to the classical $l=m+1$ Rossby modes, the HHS22 modes are very sensitive to the background density stratification.
This is because the HHS22 modes are essentially non-toroidal.
They involve substantial $z$-vortical motion near the equator and the compressional $\beta$-effect plays a significant role.
The Sun's differential rotation is also found to affect the HHS22 modes by introducing the viscous critical layers, leading to a confinement of the HHS22 modes near the base of the convection zone at large azimuthal orders $m \gtrsim 10$ (Fig.~\ref{fig:vrms_Ekin}).

We further examined possible effects of the background superadiabaticity $\delta$ on the HHS22 modes.
In this study, we used a highly-simplified radial function of $\delta(r)$ which changes from a close-to-adiabatic value in the bulk of the convection zone ($r \lesssim 0.95R_{\odot}$) to a strongly superadiabatic value near the top surface ($0.95R_{\odot} \lesssim r  \leq 0.985R_{\odot}$).
Surprisingly, the strong superadiabaticities near the top surface $\delta_{\mathrm{sf}}$ do not affect the properties of the HHS22 modes such as their propagation frequencies and the radial vorticity eigenfunction at the surface (Fig.~\ref{fig:ad_supsf}).
In contrast, we found that their dispersion relation is quite sensitive to a small variation in the bulk superadiabaticity $\delta_{\mathrm{cz}}$.
This difference can be attributed to the fact that the radial motions of the HHS22 modes dominantly exists in the lower convection zone as a consequence of the mode confinement towards the base by the differential rotation (Figs.~\ref{fig:eigenfunc_lin}c and \ref{fig:disp_growth_supsf}c).
We showed that, in order to explain the observed frequencies of the HHS22 modes, the bulk of the convection zone needs to be weakly subadiabatic, i.e., $-5\times 10^{-7} \lesssim \delta_{\mathrm{cz}} \lesssim -2.5\times 10^{-7}$.
This constraint is consistent with but tighter than the previous constraint derived by \citet[][]{gizon2021} using the $m=1$ high-latitude inertial mode ($\delta < 2\times 10^{-7}$).

Our result suggests that the Sun's bulk convection zone is likely much less superadiabatic than typically thought and possibly be even subadiabatic.
This unconventional conclusion needs to be tested by models with more realistic effects included.
For instance, we ignored the effects of magnetic field in this study, which are known to affect the properties of the Rossby modes \citep[e.g.,][]{zaqarashvili2021}.
Moreover, we still cannot rule out the possibility that the above conclusion could be affected by an inclusion of the realistic solar photosphere (above $0.985R_{\odot}$) with a very strong superadiabaticity on the order of $\delta \approx \mathcal{O}(10^{-1})$.
We also note that the spatially uniform $\delta$ below the strongly superadiabatic near-surface layer might be over-simplifying.
A further parameter survey on various radial profiles of $\delta(r)$ will be required (Dey et al. in prep.).

Recent direct numerical simulations of solar convection have reported a formation of the weakly subadiabatic layer near the base of the convection zone as a consequence of the non-local convective heat transport \citep[][]{kapyla2017,hotta2017,bekki2017b,karak2018,nelson2018,hotta2022,kapyla2023} (see also Appendix~\ref{app:nonlinear}).
Our study implies that the HHS22 modes are likely gravito-inertial modes originating from this weakly subadiabatic lower convection zone.
A further investigation on the HHS22 modes will be desired both from observational and theoretical perspectives to better understand this weakly subadiabatic layer, as it might help us to explain the Sun's anomalously weak convective power at large spatial scales \citep[e.g.,][]{hanasoge2012} and to resolve the Sun's convective conundrum \citep[e.g.,][]{omara2016,hotta2023}.

\begin{acknowledgements}
The author thanks an anonymous referee for constructive comments.
Y.~B. is also grateful to P.~Dey, R.~Cameron and L~.Gizon for helpful discussions and useful comments on the initial manuscript.
Y.~B. acknowledges a support from ERC Synergy Grant WHOLE SUN 810218.
All the numerical computations were performed at GWDG and the Max-Planck supercomputer RZG in Garching. 
\end{acknowledgements}

\bibliographystyle{aa} 
\bibliography{ref} 

\begin{appendix} 

\section{Nonlinear simulation of solar rotating convection} \label{app:nonlinear}

In this appendix, we report an additional analysis of the fully-nonlinear numerical simulations of rotating convection by \citet[][]{bekki2022b}.
We must note that these direct numerical simulations tend to show several inconsistencies with the observations regarding the profiles of the large-scale mean flows and the surface velocity power spectra \citep[e.g.,][]{miesch2008,hanasoge2012}.
This yet-unsolved problem is known as the \textit{convective conundrum} \citep[e.g.,][]{omara2016,hotta2023}.
The purpose of this appendix is, therefore, not to reproduce the observations in a self-consistent manner but only to report that the HHS22 modes can be found in the nonlinear simulations of \citet[][]{bekki2022b}.

\begin{figure*}
\begin{center}
\includegraphics[width=0.95\linewidth]{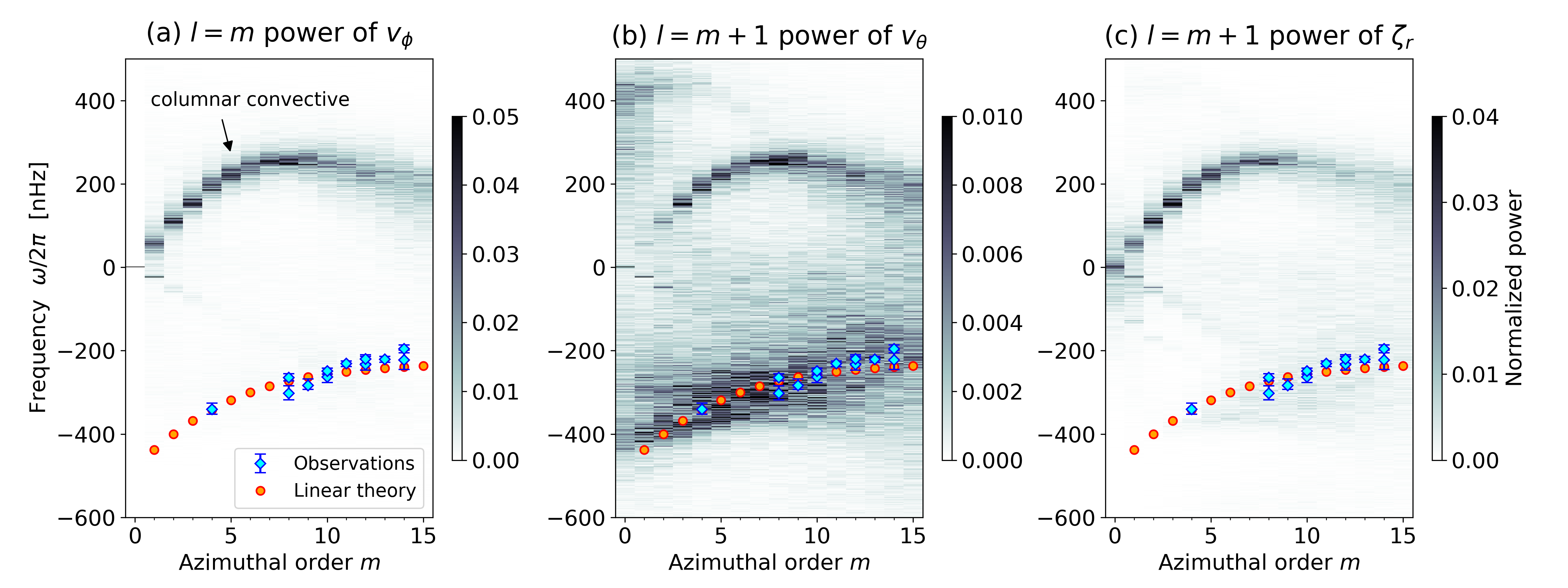}
\caption{
Power spectra from the nonlinear rotating convection simulation of (a) the $l=m$ component of longitudinal velocity $v_{\phi}$, (b) the $l=m+1$ component of latitudinal velocity $v_{\theta}$, and (c) the $l=m+1$ component of radial vorticity $\zeta_{r}$ near the surface $r=0.95R_{\odot}$.
The spectra are computed in the Carrington frame.
The power is normalized at each $m$.
Cyan diamonds denote the observed frequencies of the HHS22 modes \citepalias[][]{hanson2022}.
Orange points show the dispersion relation of the HHS22 modes obtained from the linear calculation (model 3-sub with $\delta_{\mathrm{cz}}=-5\times 10^{-7}$ and $\delta_{\mathrm{sf}}=10^{-3}$).
The power ridge associated with the columnar convective modes is denoted by black arrows.
}
\label{fig:power_nonlin}
\end{center}
\end{figure*}

\begin{figure}
\begin{center}
\includegraphics[width=0.75\linewidth]{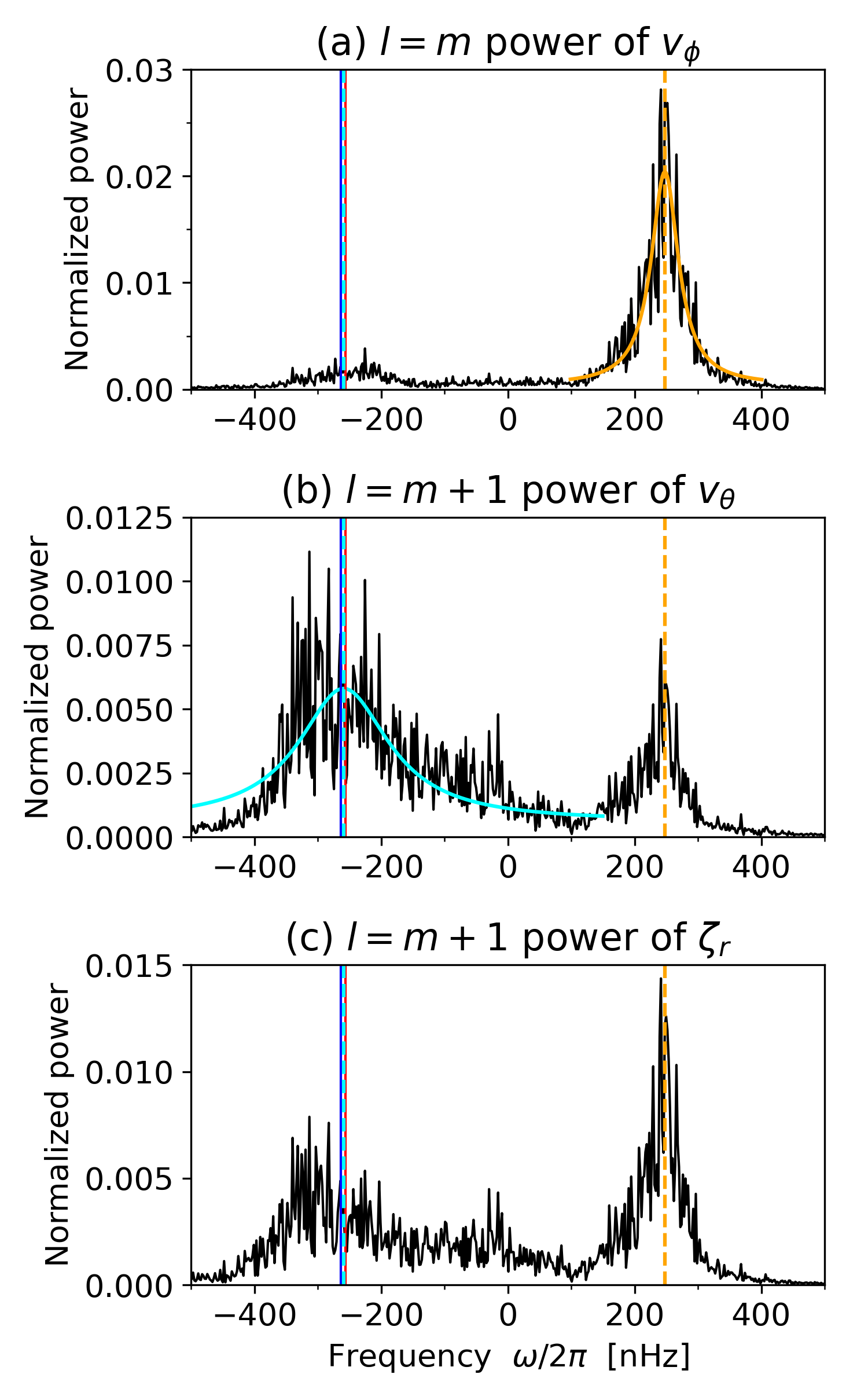}
\caption{
The same power spectra as in Fig.~\ref{fig:power_nonlin} but showing the slices at fixed azimuthal order $m=10$.
The red and blue solid lines deonte the frequencies of the $m=10$ HHS22 mode obtained from the linear calculation and measured by the observation, respectively.
The cyan solid curve in panel (b) shows the Lorentzian fit for the spectra around the HHS22 mode power peak, and the green dashed line represents the fitted peak frequency.
The same Lorentzian fitting is also performed for the columnar convective mode in panel (a) and shown by orange solid curve and dashed line.
}
\label{fig:power_m10}
\end{center}
\end{figure}

\begin{figure}
\begin{center}
\includegraphics[width=0.885\linewidth]{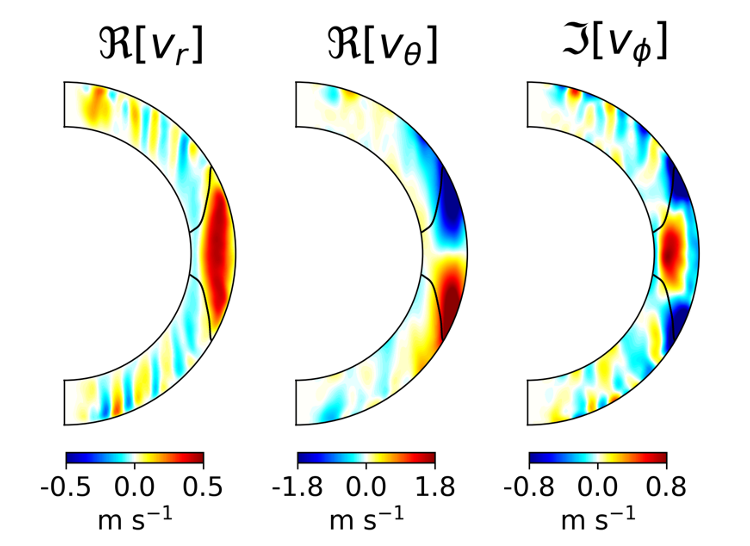}
\caption{
Velocity eigenfunctions of the $m=10$ HHS22 mode extracted from the fully-nonlinear rotating convection simulation using SVD.
}
\label{fig:eigenfunc_nonlin}
\end{center}
\end{figure}
\begin{figure}
\begin{center}
\includegraphics[width=0.9\linewidth]{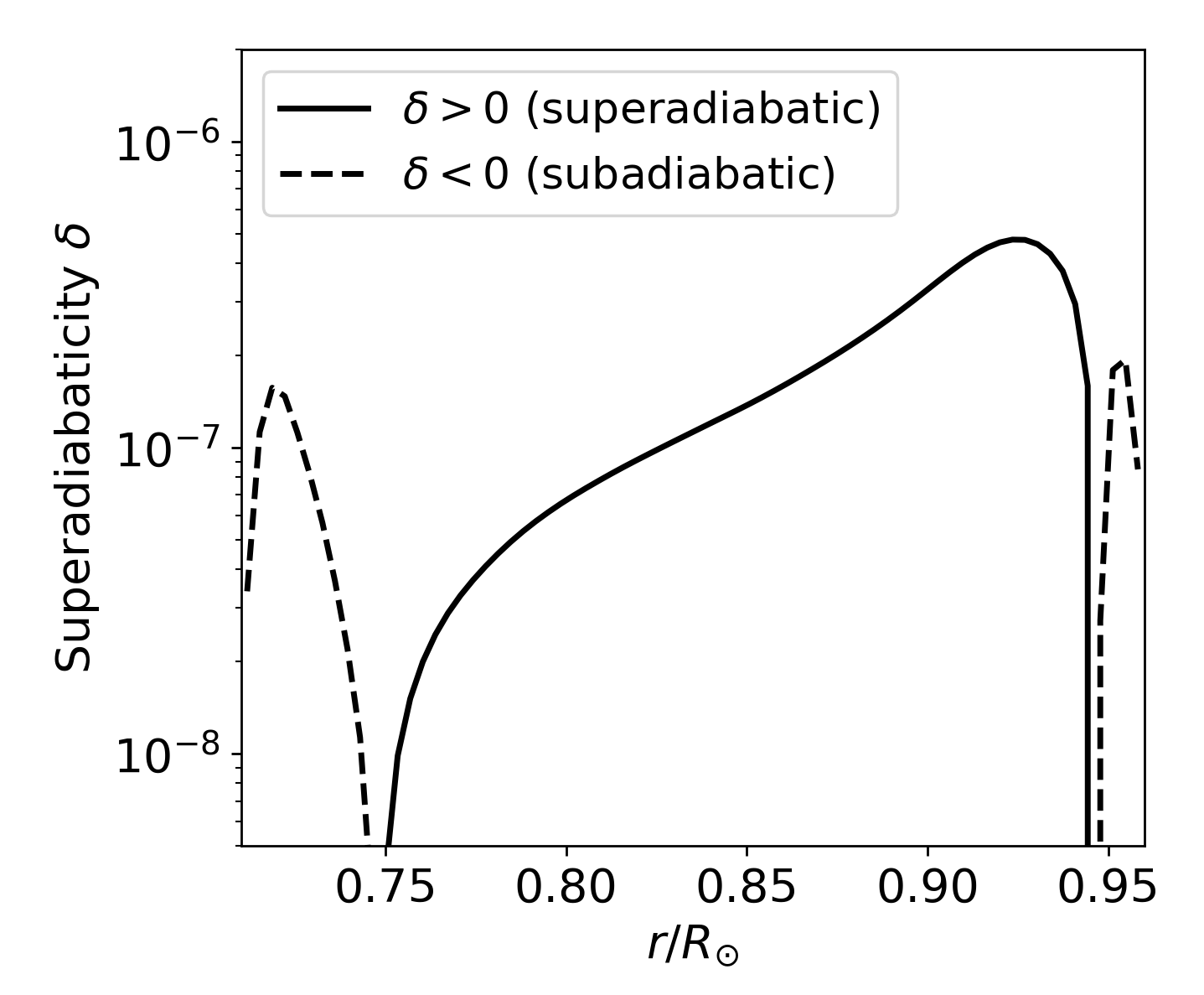}
\caption{
Radial profile of the superadiabaticity $\delta$ from the nonlinear rotating convection simulation.
Black solid and dashed curves denote the area where the mean entropy stratification is superadiabatic ($\delta>0$) and subadiabatic ($\delta<0$), respectively.
}
\label{fig:del_nonlin}
\end{center}
\end{figure}

\subsection{Method} \label{app:nonlin_method}

We use the data from the nonlinear numerical simulations of solar-like rotating convection carried out by \citet[][]{bekki2022b}.
In their simulations, a full set of compressible hydrodynamic equations are solved in a spherical shell ($0.71R_{\odot}<r<0.96R_{\odot}$) rotating at the solar rotation rate.
The luminosity is artificially reduced from the solar value by a factor of $20$ to achieve the strongly rotationally-constrained regime (low Rossby number regime) and to obtain the solar-like differential rotation with faster equator and slower poles \citep[e.g.,][]{gastine2013}.
In \citet[][]{bekki2022b}, total six simulations were carried out with the same numerical setup but with different initial random perturbations.
Each simulation has been evolved for more than $25$ solar years and we analyze the $15$-year-long data after the large-scale mean flows become statistically stationary.
The results are averaged over these six realizations to increase the signal-to-noise ratio.

\subsection{Results} \label{app:nonlin_result}

Figures~\ref{fig:power_nonlin}a--c show the near-surface power spectra of the $l=m$ component of the longitudinal velocity $v_{\phi}$, $l=m+1$ component of the latitudinal velocity $v_{\theta}$, and $l=m+1$ component of the radial vorticity $\zeta_{r}$ from the nonlinear simulations.
The spectra are computed within a Carrington frame.
Cyan diamonds denote the observed frequencies of the HHS22 modes \citepalias[][]{hanson2022} and red points show the linear dispersion relation from the model 3 with the weakly subadiabatic bulk convection zone $\delta_{\mathrm{cz}}=-5\times 10^{-7}$ which best-fits the observation (we will call this model 3-sub hereafter).
In the $l=m$ spectra of $v_{\phi}$ and in the $l=m+1$ spectra of $\zeta_{r}$, the prograde-propagating columnar convective (thermal Rossby) modes are dominantly seen \citep[][]{bekki2022b} but the HHS22 modes are not clearly visible.
On the other hand, in the $l=m+1$ spectra of $v_{\theta}$, we can see a concentration of the surface velocity power near the expected frequency range of the HHS22 modes.

\renewcommand{\arraystretch}{1.2}
\begin{table}[]
 \begin{center} 
\caption{
Properties of the HHS22 modes in our nonlinear rotating convection simulations for $1 \leq m \leq 15$.
The values in parenthesis show the results from our linear eigenmode analysis from the model 3-sub ($\delta_{\mathrm{cz}}=-5\times 10^{-7}$ and $\delta_{\mathrm{sf}}=10^{-3}$).
The frequencies are measured in the Carrington frame.
The peak frequencies and linewidths (full widths at half maxima) are obtained by Lorentzian fits.
The quantity max($v_{\mathrm{hori}}$) represents the maximum surface horizontal velocity amplitude of the mode eigenfunction extracted from the nonlinear simulation using singular-value-decomposition (SVD).
}
\small
\begin{tabular}{ccccccccccccc} 
\hline
\hline
\multirow{2}{*}{$m$} & Frequency & Linewidth & max($v_{\mathrm{hori}}$)  \\
& [nHz] & [nHz] & [m~s$^{-1}$] \\
\hline
1 & -394.3 (-437.6) & 90.3 (166.7)  & 0.4 \\
2 & -360.8 (-399.9) & 104.4 (144.7) & 0.8 \\
3 & -328.7 (-368.4) & 161.5 (124.8) & 0.7 \\
4 & -316.4 (-341.2) & 133.2 (108.5) & 0.4 \\
5 & -317.0 (-318.5) & 128.4 (96.1) & 1.9 \\
6 & -296.2 (-299.8) & 133.0 (87.3) & 0.7 \\
7 & -293.7 (-284.6) & 111.1 (81.9) & 1.9 \\
8 & -285.4 (-272.7) & 121.8 (79.0) & 1.0 \\
9 & -260.9 (-263.3) & 156.3 (78.2) & 2.1 \\
10 & -256.0 (-256.1) & 205.5 (79.3) & 2.3 \\
11 & -220.7 (-250.3) & 225.4 (81.9) & 1.1 \\
12 & -216.7 (-245.6) & 237.8 (85.6) & 3.2 \\
13 & -198.8 (-241.8) & 158.2 (89.9) & 1.3 \\
14 & -195.0 (-238.8) & 181.7 (94.7) & 3.4 \\
15 & -194.7 (-236.3) & 151.5 (99.8) & 3.5 \\
\hline
\end{tabular}
\label{table:2}
\end{center}
\end{table}
\renewcommand{\arraystretch}{1.0}

Figure~\ref{fig:power_m10} shows the same power spectra as Fig.~\ref{fig:power_nonlin} but at the fixed azimuthal order $m=10$.
It is shown that the columnar convective mode has a strong power peak at $\omega/2\pi \approx 240$ nHz with narrow linewidth of about $50$ nHz which can be seen in all the spectra (Figs.~\ref{fig:power_m10}a--c).
In the $l=m+1$ spectrum of $v_{\theta}$, another strong power concentration is seen at around $\omega/2\pi \approx -200$ nHz aside from that of the columnar convective modes, which we identified as a HHS22 mode.
To further characterize the mode properties, we perform a Lorentzian fit to this power spectrum as shown by a cyan curve in Fig.~\ref{fig:power_m10}b.
The measured peak frequencies and linewidths of the HHS22 modes in the nonlinear simulations are reported in Table~\ref{table:2} for $1 \leq m \leq 15$.
For comparison, we also report the values computed from the linear model 3-sub.
It is found that, in our nonlinear simulations, the HHS22 modes have very broad linewidths of about $100-200$ nHz, which are about twice larger than those from the linear analysis.
This suggests that they are very short-lived in the nonlinear simulations.

In order to extract the spatial eigenfunctions of the HHS22 modes from the simulation data, we apply the singular-value decomposition (SVD) method to the $l=m+1$ power spectrum of $v_{\theta}$ for each $m$ separately.
For further details on the SVD eigenmode extraction, see \citet[][\S~3.2]{bekki2022b}.
Figure~\ref{fig:eigenfunc_nonlin} shows the extracted velocity eigenfunctions of the HHS22 mode at $m=10$.
Qualitatively, a good similarity can be confirmed in their overall spatial pattern of the velocity eigenfunctions with the linear model (Figs.~\ref{fig:eigenfunc_lin}c and ~\ref{fig:disp_growth_supsf}c).
However, there exist slight differences such as the small-scale noise at high latitudes and the number of radial nodes at the equator.
These can be attributed to the effects of stochastic turbulent convection and the difference in the latitudinal differential rotation profiles.
In fact, the critical layers are formed closer to the equator (and even in the middle convection zone at high $m$) in the nonlinear simulations.

As reported in Table~\ref{table:2}, typical velocity amplitudes of the HHS22 modes are $v_{\theta} \approx 0.5-3$ m~s$^{-1}$ in the simulations, which are much weaker than those of the columnar convective modes but are comparable to those of the equatorial Rossby modes \citep[][]{bekki2022b}.
It is speculated that the HHS22 modes are excited and damped by turbulent convection similarly to the equatorial Rossby modes \citep[][]{philidet2023}.

Lastly, we show in Fig.~\ref{fig:del_nonlin} the radial profile of the superadiabaticity $\delta$ from the nonlinear rotating convection simulation.
In the nonlinear simulation, the entropy stratification is subadiabatic near the base ($r \lesssim 0.75R_{\odot}$) and superadiabatic in the upper bulk of the convection zone ($0.75R_{\odot} \lesssim r \lesssim 0.95R_{\odot}$).
The formation of the weakly subadiabatic layer near the base is a direct consequence of the non-local convective heat transport \citep[e.g.,][]{kapyla2017,bekki2017b,karak2018}.
In the nonlinear simulation of \citet[][]{bekki2022b}, the formation of this weakly subadiabatic layer is insignificant, leading to the radially-averaged superadiabaticity value of $\delta_{\mathrm{mean}}=1.2\times 10^{-7}$.
However, recent magnetohydrodynamic simulations suggest that the subadiabatic layer tends to be enhanced and extended to upper convection zone as the numerical resolution is increased and the small-scale dynamo is better resolved \citep[][]{hotta2022}.
In the real Sun, the mean entropy stratification is expected to be more subadiabatic than typically simulated likely because of the very efficient small-scale dynamo effect.

\end{appendix}

\end{document}